\documentclass[aps,12pt,final,notitlepage,oneside,onecolumn,
nobibnotes,superscriptaddress,noshowpacs,centertags]{revtex4}
\usepackage{graphicx}
\usepackage{caption2}
\begin{document}
\title{Escape distribution for an inclined billiard}
\author{Alan Roy}
\email{A.A.Roy@soton.ac.uk}
\affiliation{University of Southampton, Electrical Power Engineering Research Group,ECS, Highfield, Southampton, UK}
\author{Nikolaos Georgakarakos}
\email{georgakarakos@hotmail.com}
\affiliation{128 V. Olgas str., Thessaloniki 54645, Greece}

\begin{abstract}
H${\acute{e}}$non [8] used an inclined billiard to investigate
aspects of chaotic scattering which occur in satellite encounters
and in other situations. His model consisted of a piecewise mapping
which described the motion of a point particle bouncing
elastically on two disks. A one parameter family of orbits, named
h-orbits, was obtained by starting the particle at rest from a
given height.  We obtain an analytical expression for the escape
 distribution of the h-orbits, which is also compared with results
from numerical simulations.  Finally, some discussion is made about 
possible applications of the h-orbits in connection with Hill's problem.\\
\end{abstract}
\maketitle
\indent{MSC2010 numbers: 37D45 \hspace{0.1cm} ${\cdot}$ \hspace{0.1cm} 
37D50 \hspace{0.1cm} ${\cdot}$ \hspace{0.1cm} 70B05 \hspace{0.1cm} ${\cdot}$  \hspace{0.1cm} 70F07}

\section{INTRODUCTION}

Chaotic scattering is a phenomenon which appears in many scientific fields, 
such as astronomy, electromagnetism, statistical mechanics,
chemistry, quantum mechanics, just to mention a few.  Two objects that 
are separated initially by some distance, come closer, interact with each other
for a while and then they separate again.

More specifically, in astronomy, chaotic scattering can be found in many 
situations, such as, for example, in the formation and evolution of the outer
solar system (e.g. [1]),  in planetary rings (e.g. [11]), in
exosolar planetary systems (e.g. [4]), in the dynamical evolution
of globular clusters (e.g. [3]).  Regarding the last situation,
stars are expected to escape from a globular cluster due to a
variety of dynamical processes, such as two body relaxation or an
external tidal field. This theoretical prediction has also been
confirmed by observations (e.g. [9]). It is important
to know whether a star will escape from a cluster and also the
timescale on which that will happen as it affects the scaling of the
N-body simulation results (normally, the number N of stars used by a simulation 
is less than the number of stars of the cluster under study) and  reduce the computational 
effort required to perform a simulation.  Also, the timescale of escape is important
when one uses Monte Carlo models in order to study the dynamical evolution of globular
clusters.  Further reasons regarding the 
importance of escape in globular clusters are discussed in [6].

A star can escape from a cluster when its energy exceeds some
critical threshold but escape may not occur immediately.
Numerical simulations show that it may take a long time for a star
to escape or it may not even escape at all (escape on a timescale comparable with the age of
the universe falls into this category too). The assumption of
rapid escape referred to in [5] and [2]
is therefore a significant source of error for theoretical
predictions regarding the evolution of globular clusters. Interest has
therefore been generated in the dynamics underpinning the escape
mechanism.

Petit and H${\acute{e}}$non [10] investigated numerically
the motion of two satellites around a planet.  However, 
certain difficulties arose in their study and in order to overcome
those difficulties, H${\acute{e}}$non [8] devised a model problem which
demonstrated similar behaviour to the satellite problem but it was
easier to study. That model consisted of an inclined billiard, i.e. a point 
particle bouncing on two fixed disks and a two dimensional mapping was
used to describe the dynamics of the model. 

In the present paper, we extend the results obtained in [8] by deriving an 
analytical expression for the escape distribution of a particular 
class of orbits, the so called h-orbits.  The structure of the paper is roughly as 
follows: first, we give a description of the model developed by 
H${\acute{e}}$non (Sec. 2).  Then we obtain an analytical expression for 
the escape distribution of the h-orbits and 
the formula is compared with results from numerical simulations (Sec. 3).  
Next, there is some discussion about the billiard model and  Hill's 
problem, which originally inspired H${\acute{e}}$non  
to devise the inclined billiard model (Sec. 4).  Finally, the last section of
the paper gives a brief summary.

\section{INCLINED BILLIARD MODEL}

H${\acute{e}}$non's model is described as follows: a point
particle moves in the (X,Y) plane and bounces elastically on two
fixed disks of radius ${r}$ whose centres are located at
${(-1,-r)}$ and ${(1,-r)}$.  In addition, the particle is subject
to a constant acceleration ${g}$ in the negative ${Y}$ direction.
For reasons of simplicity, ${r}$ is considered to be large, an
assumption which means that the two disks overlap.

A one parameter family of orbits, the so called h-orbits, 
is defined by assuming that the
particle is dropped from rest at ${(h,Y_{0})}$.  ${Y_{0}}$ is a
positive constant which fixes the energy and ${h}$ is a variable.
For those orbits, there are intervals of ${h}$ in which the orbit
changes continuously and there are also critical values of ${h}$
at which a transition in behaviour is observed. For ${h=1}$ the
particle bounces ad infinitum on the right disk, while for
${h=-1}$ the particle bounces ad infinitum on the left disk.  
If the particle is
dropped either to the left of the left disk or to the right of the
right disk, then the particle escapes and never returns.  If the
particle is dropped either to the right of the left disk or to the
left of the right disk, then the particle exhibits a more complex
behaviour, bouncing from one disk to another. Hence, the values ${h=\pm 1}$ are
associated with a transition.  In general, transitions  are
present at values of ${h}$ which lead to solutions which approach
asymptotically one of the ${h=\pm 1}$  orbits.

A convenient way of dealing with the problem is to introduce a two
dimensional explicit mapping, which is equivalent to studying a
Hamiltonian system with two degrees of freedom.  Details about the
mapping and its properties can be found in [8].  If ${(X_{j},Y_{j})}$ 
is the position of the particle on the
${j}$'th rebound, then the two dimensional mapping, keeping the
notation of [8], is the following:

\begin{eqnarray}
X_{j+1} & = &
X_{j}\cosh{\phi}+w_{j}\sinh{\phi}-s_{j}(\cosh{\phi}-1)\label{eq1}\\
w_{j+1} & = &
X_{j}\sinh{\phi}+w_{j}\cosh{\phi}-(s_{j}\cosh{\phi}+s_{j+1})\tanh{\frac{\phi}{2}}\label{eq2}.
\end{eqnarray}
${w_{j}}$ is defined by
\begin{displaymath}
W_{j}=\frac{w_{j}g}{2\sqrt{2E}}\sinh{\phi},
\end{displaymath}
where ${W_{j}}$ is the transverse velocity, ${E}$ is the total
energy and ${\phi}$ is a dimensionless parameter defined by
\begin{displaymath}
\cosh{\phi}=1+\frac{4E}{gr},\hspace{1cm}
\sinh{\phi}=\sqrt{\frac{4E}{gr}(2+\frac{4E}{gr})}. 
\end{displaymath}
Finally, ${s_{j}=signX_{j}}$.  If the mapping is applied repeatedly, we obtain:
\begin{eqnarray}
\label{eq3} X_{n} &=&
\frac{h\cosh{[(n-\frac{1}{2})\phi]}}{\cosh{\frac{\phi}{2}}}-2\tanh{\frac{\phi}{2}}
\sum^{n-1}_{j=1}s_{j}\sinh{[(n-j)\phi]},\\
w_{n}
&=&\frac{h\sinh{[(n-\frac{1}{2})\phi]}}{\cosh{\frac{\phi}{2}}}-2\tanh{\frac{\phi}{2}}
\sum^{n-1}_{j=1}s_{j}\sinh{[(n-j)\phi]}-s_{n}\tanh{\frac{\phi}{2}}\nonumber.
\end{eqnarray}

A given h-orbit is associated to a sequence of binary digits (0
or 1 depending on whether the particle bounces on the left or right disk respectively)
and the corresponding to that sequence number A, with ${0\leq A \leq 1}$. 
An h-orbit changes continuously within an interval of continuity.
In that interval, the sequence of rebounds remains the same
throughout the whole of this interval.  This means that the value
of A is constant.  If A is plotted as a function of h (for a fixed
value of ${\phi}$) a fractal picture is generated.  This has the
appearance of the Devil's staircase, which consinsts of an
infinite number of horizontal bars.  Each bar corresponds
to an interval of continuity, meaning that when a particle starts
within that interval, it will follow the same orbit in terms of
which disk it bounces on. Such an example is given in fig.1.

\section{THE ESCAPE DISTRIBUTION}
\subsection{Analytical derivation}
After the brief introduction to H${\acute{e}}$non's model, we are going to obtain an expression for the number of bounces it takes
for an h-orbit to escape from the system.  The derivation will apply
for  ${e^{\phi}\leq \frac{1}{3}}$ or ${e^{\phi}\geq3}$. When
${\frac{1}{3}<e^{\phi}< 3}$, the Devil's staircase is not
continuous any longer and gaps appear between the horizontal bars,
i.e. there exist values of ${A}$ to which no h-orbits correspond.
Also, many of the assumptions made about the h-orbits do not hold.
More details about that situation can be found in
[8], although, for reasons not known to us,
H${\acute{e}}$non only refers to properties regarding
${e^{\phi}=3}$. The mapping given by Eqs (\ref{eq1}) and
(\ref{eq2}) has two fixed points (${X=-1, w=0}$ and ${X=1, w=0}$),
each one with eigenvalues ${e^{\phi}}$ and ${e^{-\phi}}$, which
suggests that when ${e^{\phi}\geq 3}$, we get at the same time
${e^{-\phi}\leq \frac{1}{3}}$.  For ${e^{\phi}=1}$, all the points
of the mapping are fixed points.  However, as stated in
[8], it is not clear whether the
peculiarities of the ${\frac{1}{3}<e^{\phi}< 3}$ case have general
relevance to the problem of chaotic scattering.

The time to escape, for a given value of h, is the smallest
integer ${k}$ such that ${|X_{n}|>1}$ for all ${n>k}$. In
particular, we would like to know which values of h, i.e. which
sub-intervals of ${(-1,1)}$, correspond to orbits which escape
after at least ${k}$ bounces within the constrained region
(${-1\leq X \leq1}$).  By summing up the lengths of these
sub-intervals it is possible to produce an escape distribution.

According to [8], at each end of every horizontal bar there
are left and right asymptotic orbits corresponding to 
\begin{equation}
\label{eq4}
h_{+}=(e^{\phi}-1)\sum^{p-1}_{j=1}e^{-j\phi}s_{j}+(e^{\phi}-2)e^{-p\phi}
\end{equation}
and
\begin{displaymath}
h_{-}=(e^{\phi}-1)\sum^{p-1}_{j=1}e^{-j\phi}s_{j}-(e^{\phi}-2)e^{-p\phi}
\end{displaymath}
respectively, where  ${p}$ is associated with each horizontal bar (p-bar),
indicating the number of digits which occurred in the binary sequence
before the repetition of either zeros or ones, i.e. before the particle
escapes over the left or right disk. Those orbits take infinitely 
many
bounces to escape.  If we move away from those extremes and
consider ${h=h_{+}-\delta}$ or ${h=h_{-}+\delta}$, we find orbits
which escape after a finite number of bounces.  The smaller ${\delta}$
is, the longer it takes for an orbit to escape (more precisely,
${k \rightarrow \infty \hspace{0.1cm}as \hspace{0.1cm} \delta
\rightarrow 0}$).  For values of h in the interval
${(h_{-},h_{+})}$, for a given p-bar, the number of bounces never
falls below p.

In order to find those intervals of ${h}$ which correspond to
orbits that remain bound for at least ${k}$ bounces, we require
to sum together the full lengths of the horizontal bars for which
${p\geq k}$.  Only partial contributions will be required from the
bars with ${p<k}$.  This is because some values of ${h}$ near the
centre of these bars lead to orbits which escape too rapidly
(under ${k}$ bounces).  The whole
calculation may be performed in the following three steps:\\
i) calculating the partial contribution from a bar with ${p<k}$ by
finding an expression for the time to escape, ${k}$, as a function
of ${\delta}$.  This expression may be inverted to yield
${\delta}$ as a function of ${k}$.  We refer to ${\delta_{r}}$ or
${\delta_{l}}$ depending on which end is being considered. The
total partial contribution from each bar is given by
${\delta_{r}+\delta_{l}}$.\\
ii) summing together the partial contributions (of which there are
${2^{p-1}}$) from all the p-bars for each ${p<k}$.\\
iii) considering the contributions from all bars with ${p\geq k}$
(the length of a bar is ${h_{+}-h_{-}}$).

We start with the first step, calculating an expression for
${k(\delta)}$ by considering the values of h which lead to left
and right asymptotic orbits.  Substituting
Eq. (\ref{eq4}) into Eq. (\ref{eq3}), we obtain
\begin{eqnarray}
X_{n}&=&\left[(e^{\phi}-1)\sum^{p-1}_{j=1}e^{-j\phi}s_{j}+(e^{\phi}-2)e^{-p\phi}\right]
\frac{\cosh{[(n-\frac{1}{2})\phi]}}{\cosh{\frac{\phi}{2}}}-\nonumber\\
&-& 2\tanh{\frac{\phi}{2}}\sum^{n-1}_{j=1}s_{j}\sinh{[(n-j)\phi]}\nonumber.
\end{eqnarray}
If we expand the above equation (keep in mind that ${s_{p}=1}$ and
${s_{j}=-1}$ for all ${j \geq p+1}$ since we deal with a left
escaping orbit), we end up with
\begin{displaymath}
X_{n}=-1+e^{-n\phi}L_{l}, \hspace{0.7cm} n\geq p+1
\end{displaymath}
where
\begin{displaymath}
L_{l}=\frac{e^{\phi}(e^{\phi}-2)e^{-p\phi}}
{e^{\phi}+1}+\frac{e^{(p+1)\phi}}{e^{\phi}+1}+\tanh{\frac{\phi}{2}}\left[e^{p\phi}+e^{\phi}
\sum^{p-1}_{j=1}e^{-j\phi}s_{j}+\sum^{p-1}_{j=1}e^{j\phi}s_{j}\right].
\end{displaymath}

Now suppose that ${h=h_{+}-\delta_{l}}$ and that the ${s_{j}}$'s
are the same as in Eq. (\ref{eq4}), i.e we are looking at the
same bar.  Then, from (\ref{eq3}), we get
\begin{eqnarray}
X^{'}_{n}&
=&(h_{+}-\delta_{l})\frac{\cosh{[(n-\frac{1}{2})\phi]}}{\cosh{
\frac{\phi}{2}}}-2\tanh{\frac{\phi}{2}}\sum^{n-1}_{j=1}s_{j}\sinh{[(n-j)\phi]}=\nonumber\\
&=&
X_{n}-\delta_{l}\frac{\cosh{[(n-\frac{1}{2})\phi]}}{\cosh{\frac{\phi}{2}}}=
-1+e^{-n\phi}L_{l}-\delta_{l}\frac{\cosh{[(n-\frac{1}{2})\phi]}}{\cosh{\frac{\phi}{2}}}.
\label{eq5}
\end{eqnarray}
As we are interested in the number of bounces ${k}$ before the
particle escapes over the left disk,  Eq. (\ref{eq5})
yields (${X^{'}_{k}=-1}$)
\begin{displaymath}
\delta_{l}=\frac{2e^{\frac{\phi}{2}}\cosh{\frac{\phi}{2}}L_{l}}{e^{2k\phi}+e^{\phi}}=
\frac{(e^{\phi}+1)L_{l}}{e^{2k\phi}+e^{\phi}}.
\end{displaymath}
Following the above approach, we can obtain a similar expression
for right asymptotic orbits.  In this case, we find that
\begin{displaymath}
X_{n}=1-e^{-n\phi}L_{r}, \hspace{0.7cm} n\geq p+1
\end{displaymath}
where
\begin{displaymath}
L_{r}=\frac{e^{\phi}(e^{\phi}-2)e^{-p\phi}}
{e^{\phi}+1}+\frac{e^{(p+1)\phi}}{e^{\phi}+1}+\tanh{\frac{\phi}{2}}
\left[e^{p\phi}-e^{\phi}\sum^{p-1}_{j=1}e^{-j\phi}s_{j}-\sum^{p-1}_{j=1}e^{j\phi}s_{j}\right]
\end{displaymath}
and eventually,
\begin{displaymath}
\delta_{r}=\frac{2e^{\frac{\phi}{2}}\cosh{\frac{\phi}{2}}L_{r}}{e^{2k\phi}+e^{\phi}}=
\frac{(e^{\phi}+1)L_{r}}{e^{2k\phi}+e^{\phi}}.
\end{displaymath}

Now, we continue with the second step, i.e. summing together all the
partial contributions.  In order to do that, we are going to use
two subscripts.  The first subscript is ${p}$, which was defined
earlier, while the second subscript ${q}$
distinguishes between bars of the same p-value.  In the case of 
bars with ${p<k}$, we must evaluate
\begin{equation}
\label{eq6}
\sum^{k-1}_{p=1}\sum^{2^{p-1}}_{q=1}[(\delta_{l})_{pq}+(\delta_{r})_{pq}],
\end{equation}
where
\begin{displaymath}
(\delta_{l})_{pq}=\frac{2e^{\frac{\phi}{2}}\cosh{\frac{\phi}{2}}(L_{l})_{pq}}{e^{2k\phi}+e^{\phi}}\hspace{0.5cm}
1\leq p \leq k-1,
\end{displaymath}
\begin{displaymath}
(\delta_{r})_{pq}=\frac{2e^{\frac{\phi}{2}}\cosh{\frac{\phi}{2}}(L_{r})_{pq}}{e^{2k\phi}+e^{\phi}}\hspace{0.5cm}
1\leq p \leq k-1,
\end{displaymath}
\begin{displaymath}
(L_{l})_{pq}=\frac{e^{\phi}(e^{\phi}-2)e^{-p\phi}}
{e^{\phi}+1}+\frac{e^{(p+1)\phi}}{e^{\phi}+1}+\tanh{\frac{\phi}{2}}\left[e^{p\phi}+e^{\phi}
\sum^{p-1}_{j=1}e^{-j\phi}s_{qj}+\sum^{p-1}_{j=1}e^{j\phi}s_{qj}\right],
\end{displaymath}
and
\begin{displaymath}
(L_{r})_{pq}=\frac{e^{\phi}(e^{\phi}-2)e^{-p\phi}}
{e^{\phi}+1}+\frac{e^{(p+1)\phi}}{e^{\phi}+1}+\tanh{\frac{\phi}{2}}\left[e^{p\phi}-e^{\phi}\sum^{p-1}_{j=1}e^{-j\phi}s_{qj}
-\sum^{p-1}_{j=1}e^{j\phi}s_{qj}\right].
\end{displaymath}
When we substitute the above equations into Eq. (\ref{eq6}), the
following quantity will appear:
\begin{eqnarray}
\sum^{2^{p-1}}_{q=1}(L_{l})_{pq}&=&
2^{p-1}\left[\frac{e^{\phi}(e^{\phi}-2)e^{-p\phi}}
{e^{\phi}+1}+\frac{e^{(p+1)\phi}}{e^{\phi}+1}+e^{p\phi}\tanh{\frac{\phi}{2}}\right]+\nonumber\\
&+&
\sum^{2^{p-1}}_{q=1}\sum^{p-1}_{j=1}(e^{\phi}e^{-j\phi}+e^{j\phi})s_{qj}\tanh{\frac{\phi}{2}}.\nonumber
\end{eqnarray}
The same expression arises for ${(L_{r})_{pq}}$, except a minus
sign before the double sum. The double summation term is zero
because, for a given ${j}$, ${s_{qj}=\pm 1}$ in equal numbers of
bars. Hence:
\begin{equation}
\sum^{k-1}_{p=1}\sum^{2^{p-1}}_{q=1}[(\delta_{l})_{pq}+(\delta_{r})_{pq}]=\frac{e^{\phi}+1}{e^{2k\phi}+e^{\phi}}
\sum^{k-1}_{p=1}\sum^{2^{p-1}}_{q=1}[(L_{l})_{pq}+(L_{r})_{pq}]=\frac{2^{k}[e^{(k-1)\phi}-e^{-(k-1)\phi}]}{1+ e^{(2k-1)\phi}}.
\label{eq7}
\end{equation}

Finally, the contribution from all bars with ${p\geq k}$ is
\begin{equation}
\label{eq8}\sum^{\infty}_{p=k}2^{p-1}(h_{+}-h_{-})=\sum^{\infty}_{p=k}
2^{p-1}2(e^{\phi}-2)e^{-p\phi}=\frac{2^{k}e^{-k\phi}(e^{\phi}-2)}{1-2e^{-\phi}}.
\end{equation}
Thus, ${Esc(k)}$, the fraction of h-orbits in the interval
${(-1,1)}$ which escape after at least k bounces, is found by
adding Eq. (\ref{eq7}) and Eq. (\ref{eq8}) and dividing by 2:

\begin{equation}
Esc(k)=\frac{1}{2}\left\{\frac{2^{k}[e^{(k-1)\phi}-e^{-(k-1)\phi}]}{1+
e^{(2k-1)\phi}}+\frac{2^{k}e^{-k\phi}(e^{\phi}-2)}{1-2e^{-\phi}}\right\}=2^{k-1}\frac{e^{k \phi}(1+e^{\phi})}{e^{\phi}+e^{2k \phi}}.
\label{eq9}
\end{equation}

\subsection{Numerical Results}

In order to test our analytical result , we advanced Eq.
(\ref{eq1}) and (\ref{eq2}) numerically. The interval ${(-1,1)}$
was divided into N subintervals, each having a width
of ${2/N}$.  The initial conditions were
\begin{displaymath}
X_{0}=-1+k\frac{2}{N}, \hspace{0.7cm}
w_{0}=-(X_{0}-s_{0})\tanh{\frac{\phi}{2}}, \hspace{0.5cm}
k=1,2,...,N,
\end{displaymath}
with the initial values ${X_{0}}$ distributed uniformly in the
interval ${(-1,1)}$.  Using the iterative scheme given by
Eq. (\ref{eq1}) and (\ref{eq2}), $h$-orbits were evolved
forward in time and for each $h$-orbit, we found the minimum number
of iterations $k$ required to satisfy ${|X_{k}|>1}$.  An escape
distribution is built up by determining what fraction of the
$h$-orbits escaped on the first bounce, on the second bounce and
so on. The simulations were done for different values of ${\phi}$
and ${N}$. The value of $N$ was even increased to ${N=5000000}$ to
ensure that the results were independent of the number of
subdivisions.  Fig.2 shows the escape distribution for
different values of ${\phi}$, both analytical [Eq. (\ref{eq9})] and numerical .

The numerical results are in excellent agreement with those
obtained from Eq. (\ref{eq9}). There is only a small range
of ${e^{\phi}}$ for which there is some discrepancy between the
numerical  and the analytical results for some p-bars. Every horizontal bar has a
point ${h_{D}}$, which, according to [8], is given by 
\begin{displaymath}
h_{D}=\frac{2\sinh{\frac{\phi}{2}}}{\cosh{[(n-\frac{1}{2})\phi]}}\sum^{n-1}_{j=1}s_{j}\sinh{[(n-j)\phi]}
\end{displaymath}
and for  which ${X_{n}=0}$.  When
${h_{-}+\delta_{r}> h_{D}}$ or ${h_{+}-\delta_{l}<h_{D}}$, our
analytical result needs some correction, i.e. the difference in
bar length ${h_{D}-(h_{-}+\delta_{r})}$ or
${(h_{+}-\delta_{l})-h_{D}}$. For example, also visible in fig.2,
for ${e^{\phi}=3}$ and ${k=3}$ our analytical result is around
${3.5\%}$ larger than the numerical one and that happens because
our calculation overestimates the contribution of the left side of
the ${p=2}$ bar (for ${h>0}$; for ${h<0}$ is the right side of the
${p=2}$ bar that exhibits some problem). For ${e^{\phi}=3.2}$, the
error for ${k=3}$ reduces to around ${1\%}$. If the extra bar
length is subtracted from our analytical calculation, we obtain
the correct percentage. For the ${p=2}$ bar, the correction is:
\begin{displaymath}
h_{D}-(h_{-}+\delta_{r})=1-2(e^{-\phi}-e^{-2\phi})+\frac{1-2e^{\phi}-e^{2\phi}+2e^{3\phi}}{e^{\phi}+e^{6\phi}}-
\frac{1-e^{\phi}-e^{2\phi}+e^{3\phi}}{1+e^{3\phi}}.
\end{displaymath}

\section{HILL'S PROBLEM AND h-ORBITS}

As it was stated in the introduction, H${\acute{e}}$non's billiard
model was inspired by [10], which dealt with the interaction of two 
satellites around a planet.  That study was done in the context of   
the so called Hill's problem, a special case of the restricted 
three body problem where the massless particle moves in the neighbourhood 
of the secondary body.  Originally intended
as a model for the motion of the Moon around the Earth with
perturbations by the sun, with some modification it can also
serve as a simplified model of the dynamical behaviour of
escaping stars in globular clusters [7]. In that case, the
centre of the galaxy and the globular cluster play the role of the
two main bodies, while the star is treated as the massless particle.

Let us assume that the cluster moves on a circular orbit of 
radius ${R}$ around the centre of the galaxy with constant angular velocity,  
the mass of the cluster with respect to the galaxy is assumed to be small,
the mass of the star is considered to be negligible compared
to that of the cluster and all three bodies lie in the same plane.
If we consider a coordinate system that rotates with the angular
velocity of the cluster around the galaxy and with its origin
being at the cluster, Hill's equations can be written as follows [13]:
\begin{eqnarray}
\ddot{\xi}-2\dot{\eta}-3\xi & = &-\frac{\xi}{r^{3}}\label{eq10}\\
\ddot{\eta}+2\dot{\xi} & = & - \frac{\eta}{r^{3}}\label{eq11},
\end{eqnarray}
where ${\xi}$ and ${\eta}$ are the ${x}$ and ${y}$ coordinates of
the star and ${r=\sqrt{\xi^{2}+\eta^{2}}}$. Note that the above equations 
have been scaled in terms of time
and distance.  In the above mentioned coordinate system, the two
Lagrangian points of interest occur at
${(-\frac{1}{3}^{\frac{1}{3}},0)}$ and
${(\frac{1}{3}^{\frac{1}{3}},0)}$.

One recalls that in our inclined billiard model, the particle was
dropped with zero horizontal velocity onto the inclined surface
from a fixed height, i.e. with constant ${\phi}$ for various ${h}$
in the interval (-1,1).  In Hill's problem, the h-orbits
can be defined to be those with initial conditions given by
\begin{displaymath}
\xi_{0}=C,\hspace{0.5cm}\dot{\xi}_{0}=0,\hspace{0.5cm}\eta_{0}=0.
\end{displaymath}
At a fixed value of the initial Hamiltonian ${H_{in}}$ (analogous
to the condition of constant ${\phi}$ in H${\acute{e}}$non's
model), ${\dot{\eta}_{0}}$ is calculated from
\begin{displaymath}
H_{in}=\frac{1}{2}(\dot{\xi}^{2}_{0}+\dot{\eta}^{2}_{0})-\frac{3}{2}\xi^{2}_{0}-\frac{1}{|\xi_{0}|}.
\end{displaymath}

Each time the orbit intersects the surface of section ${\eta=0}$
(with ${\dot{\eta} >0}$ when ${\xi >0}$ and with
${\dot{\eta} <0}$ when ${\xi <0}$), the value of ${\xi}$
is recorded.  If ${\xi=\xi_{j}}$ on the ${j}$th intersection with
the surface of section, then the orbit may be represented
symbolically by a sequence ${\{d_{j}, \hspace{0.2cm}
j=1...\infty\}}$, where
\begin{displaymath}
d_{j}=\cases{0 \hspace{0.2cm} if \hspace{0.2cm} \xi_{j}<0\cr 1
\hspace{0.2cm}if \hspace{0.2cm}\xi_{j}>0. \cr}
\end{displaymath}
If the orbit escapes through the right Lagrangian point after
${k}$ intersections with the surface of section, then
${d_{j}=1, \hspace{0.2cm}j>k}$.  Similarly, if the orbit escapes 
through the left Lagrangian point
after ${k}$ intersections with the surface of section, then
${d_{j}=0, \hspace{0.2cm}j>k}$.
Hence a real number can be attached to a given orbit in a similar way as it was
done for the inclined billiard model. 

Fig.3 is an example of an h-orbit in Hill's problem, exhibiting similar 
features to those found in fig. 1.

\section{SUMMARY}

H${\acute{e}}$non [8] used an inclined billiard in order to
investigate the  phenomenon of chaotic scattering. A point
particle, initially at rest, fell from a certain height
and bounced elastically off the surface of two disks. When the
particle bounced beyond the top of either disk, it was considered
to have escaped. By assigning a number ${A}$ to the orbit of the
particle, we were able to plot ${A(h)}$, where ${h}$ (${-1 \leq h
\leq 1}$) was the initial value of the x-coordinate of the
particle. For certain values of the parameter ${\phi}$, which was
related to the particle energy, the graphical representation of
${A(h)}$ had the form of the so-called Devil's staircase.  Based
on certain properties of the Devil's staircase, we were able to
derive an analytical expression for the number of particles that
escaped after at least ${k}$ bounces.

It may be possible to approach the issue of stars escaping from a cluster or other
similar situations by using the billiard model as a simplification of Hill's problem. 
Therefore, our future aim  is to investigate the possibility of connecting
the billiard model and Hill's problem in a way that can be used to describe 
a more complex situation such as for example the dynamical evolution of a star 
within a globular cluster.  The latter is far more complicated as a system than the
inclined billiard we have studied, but many times, it is possible to get a good
approximation of a complicated system by following a more simple approach to the problem.  

\section*{ACKNOWLEDGEMENTS}
The authors would like to thank Prof. D.C. Heggie for valuable comments
regarding certain aspects of this work. Alan Roy should also like to thank Prof. Jan
Sykulski for supporting his research visitor status in ECS at the University of Southampton.

\newpage
\section*{REFERENCES}

\noindent 1. Astakhov, S.A., Lee, E.A., Farrelly, D., Formation of Kuiper-belt binaries through 
\indent multiple chaotic scattering encounters with low-mass intruders, MNRAS, 2005, vol. 360,
\indent pp. 401-415.

\noindent 2. Baumgardt, H., Scaling of N-body calculations, MNRAS, 2005, vol. 325, pp. 1323-1331.

\noindent 3. Ernst, A., Just, A., Spurzem, R., Porth O., Escape from the vicinity of fractal basin 
\indent boundaries of a star cluster,  MNRAS, 2008, vol. 383,pp. 897-906.

\noindent 4. Ford, E.B., Lystad, V., Rasio, F.A., Planet-planet scattering in the upsilon Andromedae 
\indent system, Nature, 2005, vol. 434, pp. 873-876.

\noindent 5. Fukushige, T., Heggie, D.C., The time-scale of escape from star clusters, MNRAS, 2000, 
\indent vol. 318, pp. 753-761.

\noindent 6. Heggie, D.C., Mass loss from globular clusters, {\it Dynamics of Star Clusters and the Milky 
\indent Way}, S. Deiters, B. Fuchs, A. Just, R. Spurzem, R. Wielen (Eds.), San Francisco: Astron. 
\indent Soc. Pac., 2001, ASP Conf. Ser. vol. 228, pp. 29-41.

\noindent 7. Heggie, D.C., Escape in Hill's problem, {\it The restless universe, Proceedings of the
54th 
\indent Scottish Univ. Summer School in Physics}, B.A. Steves, A.J. Maciejewski (Eds.), 
Bristol: 
\indent Scottish Universities Summer School in Physics and IoP Publishing, 2001, pp. 109-128.

\noindent 8. H${\acute{e}}$non, M., Chaotic scattering modelled by an inclined billiard, Physica D, 1988, vol. 33, 
\indent pp. 132-156.

\noindent 9. Leon, S., Meylan, G., Combes, F., Tidal tails around 20 Galactic globular clusters. 
\indent Observational evidence for gravitational disk/bulge shocking,  A${\&}$A, 2000, vol. 359, pp. 
\indent 907-931.

\noindent 10. Petit, J.M., H${\acute{e}}$non, M., Satellite encounters,  Icarus, 1986, vol. 66, pp. 536-555.

\noindent 11. Petit, J.M., Chaotic scattering in planetary rings, {\it Singularities in gravitational systems},
\indent D. Benest, C. Froeschlé (Eds.), Germany: Springer, 2002, Lect. Not. Phys. vol. 590, pp. 
\indent 114-144.

\noindent 12. Press, W.H., Teukolsky, S.A., Vetterling, W.T., Flannery, B.P., Numerical Recipes In 
\indent Fortran 77. Second edition.  New York: Cambridge Univ. Press, 1996.

\noindent 13. Szebehely, V.G., Theory of orbits, New York: Academic Press, 1967.

\newpage
\begin{figure}
\includegraphics{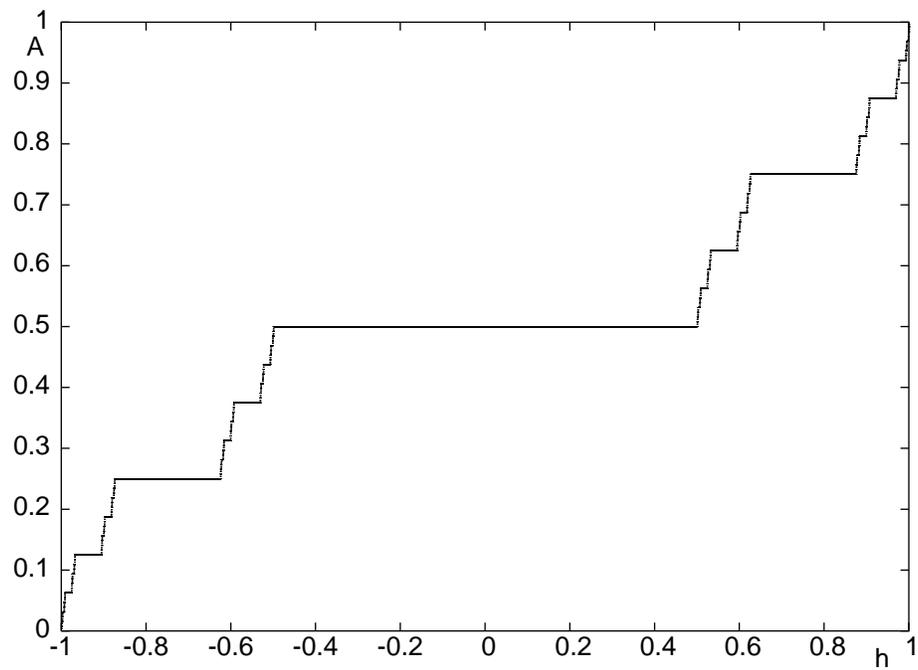}
\caption{ The Devil's staircase for ${e^{\phi}}$=4.}
\end{figure}

\newpage
\begin{figure}
\includegraphics{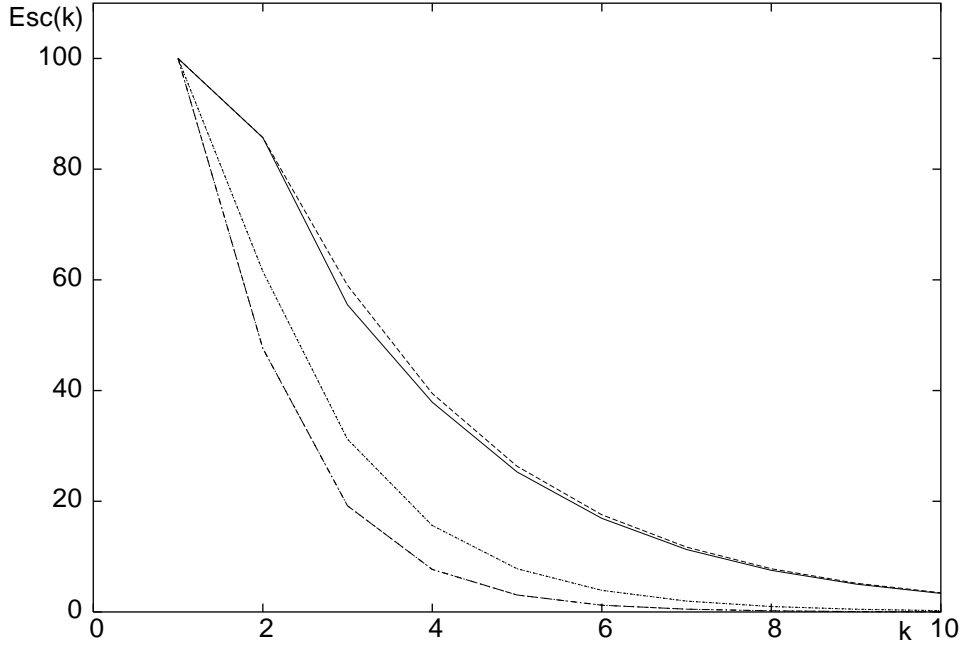}
\caption{ Percentage of particles that escape after at least ${k}$
bounces against the number of bounces ${k}$.  From right to left, the first two curves correspond to
${e^{\phi}=3}$ (the left curve comes from the simulations, while the right one is based on
Eq.(\ref{eq9})), the third one corresponds to ${e^{\phi}=4}$ and the fourth one corresponds to ${e^{\phi}=5}$.
Note that each of the third and fourth curves are actually two curves (one on top of the other one),
as the numerical and analytical results are almost identical.  Also note the small discrepancy between 
the numerical and analytical results for ${e^{\phi}=3}$.}
\end{figure}

\newpage
\begin{figure}
\includegraphics{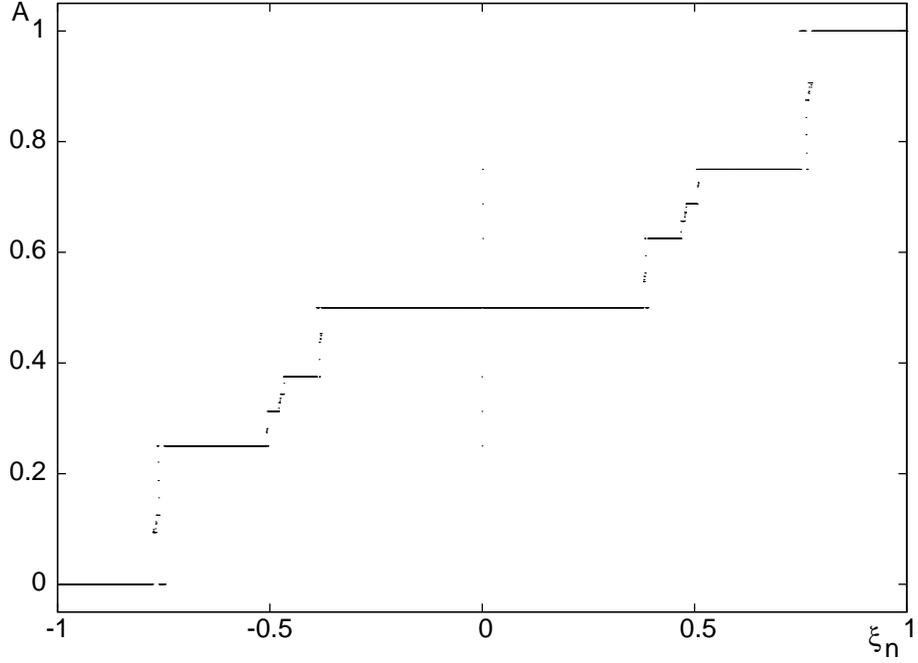}
\caption[]{ Orbital sequence number ${A}$ against ${\xi_{n}}$ in
Hill's problem, where ${\xi_{n}=\frac{1}{3}^{-\frac{1}{3}}\xi_{0}}$.  The initial value 
of the Hamiltonian is ${H_{in}=-1.9}$.  The results for Hill's problem were obtained by integrating
equations (\ref{eq10}) and (\ref{eq11}) numerically, using a
Burilsch-Stoer integrator with a variable time step [12]. 
The time of integration was ${4\pi}$ which, in our system of units, is twice the orbital period of the
cluster around the galaxy.}
\end{figure}
\vspace{10cm}

\end{document}